\documentclass[journal=jctcce,manuscript=article]{achemso}
\setkeys{acs}{articletitle=true}
\setkeys{acs}{keywords=true}
\usepackage[utf8]{inputenc}

\usepackage{graphicx}
\usepackage{amsmath}
\usepackage{amssymb}
\usepackage[hidelinks]{hyperref}
\usepackage{multirow}
\usepackage{multicol}
\usepackage{soul}
\usepackage{xcolor}
\usepackage[font=small,labelfont=bf]{caption} 
\usepackage{bm}
\usepackage[version=3]{mhchem} 

\newcommand{\vshe}{V$_\mathrm{SHE}$}
\newcommand{\vrhe}{V$_\mathrm{RHE}$}
\newcommand{\hupd}{H$_\mathrm{UPD}$}
\newcommand{\hopd}{H$_\mathrm{OPD}$}
\newcommand{\gnet}{\ensuremath{\Delta G_\mathrm{net}}}
\newcommand{\gdag}{\ensuremath{\Delta G^\ddag}}
\newcommand{\pe}{\ensuremath{\Phi_\mathrm{e}}}
\newcommand{\nne}{\ensuremath{N_\mathrm{e}}}
\newcommand{\gz}{\ensuremath{G{\sim}0}}
\newcommand{\pot}{\ensuremath{\phi}}

\newcommand{\dgrp}{\ensuremath{\Delta G_\mathrm{rxn}^\pot}}
\newcommand{\mup}[1]{\mu^\pot[\ce{#1}]}
\newcommand{\muo}[1]{\mu^\circ[\ce{#1}]}

\title{A challenge to the $\bm{\Delta G{\sim}0}$ interpretation of hydrogen evolution}
\author{Per Lindgren}
\author{Georg Kastlunger}
\author{Andrew A. Peterson}
\email{andrew\_peterson@brown.edu}
\affiliation{School of Engineering, Brown University, Providence, Rhode Island, 02912, USA}
\keywords{Hydrogen evolution, Pt(111), constant-potential DFT, Tafel slopes, current densities}

\begin{document}
\singlespace
\begin{abstract}
\noindent
Platinum is a nearly perfect catalyst for the hydrogen evolution reaction, and its high activity has conventionally been explained by its close-to-thermoneutral hydrogen binding energy (\gz).
However, many candidate non-precious metal catalysts bind hydrogen with similar strengths, but exhibit orders-of-magnitude lower activity for this reaction.
In this study, we employ electronic structure methods that allow fully potential-dependent reaction barriers to be calculated, in order to develop a complete working picture of hydrogen evolution on platinum.
Through the resulting \textit{ab initio} microkinetic models, we assess the mechanistic origins of Pt's high activity.
Surprisingly, we find that the \gz\ hydrogen atoms are kinetically inert, and that the kinetically active hydrogen atoms have $\Delta G$'s much weaker, similar to that of gold.
These on-top hydrogens have particularly low barriers, which we compare to those of gold, explaining the high reaction rates, and the exponential variations in coverages can uniquely explain Pt's strong kinetic response to the applied potential.
This explains the unique reactivity of Pt that is missed by conventional Sabatier analyses, and suggests true design criteria for non-precious alternatives.
\end{abstract}

\maketitle

\begin{multicols}{2}

\subsection*{Introduction}
Recent climate reports call for carbon emissions to be rapidly phased out in order to limit global warming's most severe impacts, calling for 50\%\ reductions by $\sim$2030 and (net) zero emissions by $\sim$2050~\cite{IPCC2018}.
While many scalable carbon-free electricity generation sources exist, the path to carbon-neutral fuels and chemicals is less obvious.
We argue that that technologies must be rapidly developed for the efficient conversion of electricity into fuels.
The most readily deployable approach is to make \ce{H2} (from \ce{H2O}) with renewable electricity, then use the \ce{H2} in existing chemical processes---such as Haber--Bosch for ammonia synthesis, Fischer--Tropsch (preceded by reverse water-gas shift) for fuel synthesis, hydrotreating for biomass upgrading, or direct use in fuel-cell vehicles.

Functionally, platinum is a nearly perfect catalyst for the hydrogen evolution reaction (HER).
In acidic media, platinum begins evolving \ce{H2} very close to the reaction's equilibrium potential, and its Tafel slope---an (inverse) measure of how strongly the reaction rate responds to changes in potential---is excellent, at $\sim$30 mV dec$^{-1}$ \cite{Fei2015, Wu2017, Jiang2017}.
It is certainly the most efficient known electrocatalyst, and perhaps also the best heterogeneous catalyst discovered to date.
However, the low abundance of Pt in the earth's crust makes scaling Pt-based processes challenging.~\cite{Vesborg2012}

Conventionally, Pt's activity has been explained by its hydrogen binding energy~\cite{Parsons1958, TRASATTI1972163}, which has been calculated to be almost perfectly thermoneutral at the equilibrium potential~\cite{Noerskov2005, Greeley2006}, maximizing the activity towards HER in accordance with the Sabatier principle.~\cite{Sabatier}
Despite extraordinary efforts to synthesize earth-abundant, inexpensive and stable catalysts based on this principle, alternate electrocatalysts with close-to-thermoneutral differential hydrogen binding energy exhibit HER activities that are orders of magnitude lower than on platinum~\cite{Hellstern2017,Hinnemann2005, Jaramillo2007, Roger2017, Chen2011, McCrory2015, Benck2014, Bonde2008, Macdonald2017, Zhang2015}.
Thus, hydrogen binding energy alone cannot explain the superior HER activity of platinum catalysts.

\begin{figure*}[t]
\begin{center}
\includegraphics[width=1.0\textwidth]{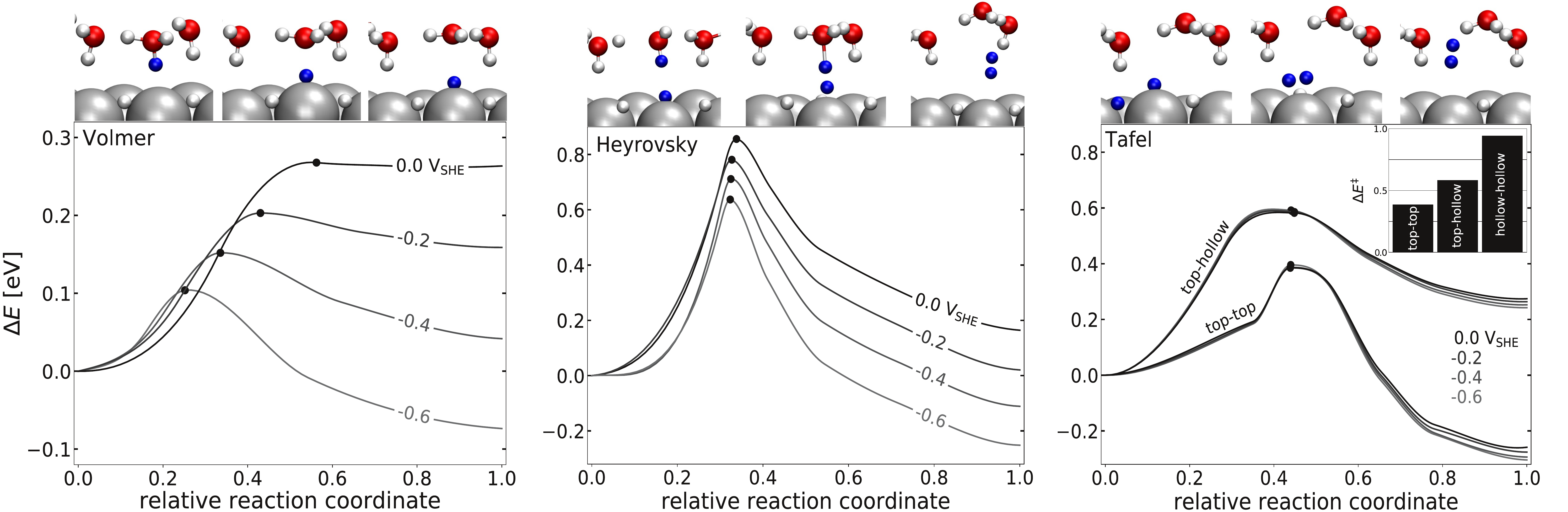}
	\caption{Minimum energy pathways of the Volmer, Heyrovsky and Tafel steps of the HER mechanism as functions of the applied potential. The insets represent the initial, transition and final states at the equilibrium potential. All reaction barriers were calculated with the SJ method at constant potential using the climbing-image nudged elastic band (CI-NEB)~\cite{Henkelman2000,CINEB} method.}
\label{fig:bands}
\end{center}
\end{figure*}

Hydrogen evolution is a two-electron reaction commonly considered in two elementary steps:
\begin{equation}\label{eq:mechanisms}
\begin{array}{ccc}
\ce{$*$ + (H+ + e^-) &->[\mathrm{Volmer}]& H$*$} \\
\ce{H$*$ + (H+ + e^-) &->[\mathrm{Heyrovsky}]& H2 + $*$} \\
\ce{2 H$*$ &->[\mathrm{Tafel}]& H2 + 2 $*$}
\end{array}
\end{equation}
\noindent
where $*$ is a catalytic site and H$*$ is a surface-bound hydrogen.
The Volmer reaction places atomic hydrogen on the catalyst surface via proton discharge from the electrolyte, coupled with electron transfer from the electrode.
\ce{H2} is liberated by either the Heyrovsky or Tafel step, as shown above.
The Volmer and Heyrovsky steps include the explicit transfer of an electron, suggesting that their elementary rate constants will have a strong potential dependence; the Tafel step is non-electrochemical and its rate constant can be expected to have negligible potential dependence, save for field effects.
Since \ce{H$*$} appears in each elementary step, it is perhaps not surprising that the hydrogen binding strength has traditionally been employed as a first descriptor of reactivity.
However, it is known that underpotential deposition of hydrogen (\hupd) also occurs on Pt electrodes.\cite{Ogasawara1994, Markovic1996, Morin1996, Conway1998,  Herrero2001, osiewicz2012}
Experimentally, this is observed in cyclic voltammograms as a capacitive event at potentials on the positive side of the equilibrium potential ($>$0 \vrhe), before \ce{H2} evolution takes off ($<$0 \vrhe).
The generally accepted microscopic view of this phenomenon based on electronic structure calculations is adsorption of H into the metallic fcc-hollow sites \cite{Greeley2005, Skulason2010, Santos2011}, creating an almost complete monolayer around the equilibrium potential of HER, before the surface begins to evolve \ce{H2}.
These \hupd\ have a free energy of binding near zero (\gz), and are conventionally used in ``volcano'' plots to explain Pt's activity.~\cite{Noerskov2005, Greeley2006}
After the hollow sites fill, the next available sites are those on top of Pt; these overpotential-deposited H's (\hopd) have a much weaker binding free energy, similar to that of Au or Ag~\cite{Santos2011}, which are relatively poor catalysts for this reaction.
Thus, there is apparent contradiction---particularly from the computational side---about the roles of these two types of bound hydrogen.~\cite{Gomez1993, Markovic1996, Santos2011}

Clearly, to describe the activity of platinum at a more sophisticated level requires knowledge of potential-dependent reaction barriers. However, while conventional density functional theory (DFT)~\cite{Hohenberg_Kohn_DFT_1964, Kohn_Sham_DFT_1965} can easily be applied to study the thermodynamics of HER, reaction barrier calculations~\cite{Skulason2010, Chan2015} often suffer from little control over potential and coverages and from the need to extrapolate to constant-workfunction behavior, making the methods computationally expensive and the interpretation of results challenging.

Introduced 15 years ago, the computational hydrogen electrode (CHE)~\cite{Norscov_Ross_JPC_2004} has revolutionized computational electrocatalysis with its computational efficiency and thermodynamic rigor.
Here, the proton--electron pair is in equilibrium with gaseous hydrogen at the equilibrium potential, and can be substituted by the energy of gaseous hydrogen. CHE has allowed for calculations of activity trends based on thermodynamic descriptors alone, but has recently been criticized for its overly simplistic connection between thermodynamic descriptors and kinetic metrics as well as its inability to correctly predict activity trends in reactions with overpotential-dependent Tafel slopes~\cite{Exner2019}. However, despite its many successes, CHE is a purely thermodynamic construct, and does not allow for reaction barrier calculations as a function of the applied potential.

The advent of grand-canonical electronic structure schemes has allowed for explicit control of applied potential in atomistic studies of electrocatalytic systems by varying the number of electrons, allowing the workfunction to be tuned relative to some reference potential \cite{Alavi_JCP_2001, Sugino_PRB_2006, Neurock_2006, Jinnouchi2008, Fang2009, Fang2010, Letchworth-Weaver2012, Head-Gordon_2016,Sundararaman_GC_2017, Kastlunger2018, Bouzid2018, Melander2019}. These methods provide a straightforward way of calculating reaction barriers at constant potential.
In particular, the solvated jellium (SJ) method~\cite{Kastlunger2018} accomplishes potential control for periodic systems by employing a counter charge in a solvated jellium slab, which localizes the excess charge on the reactive side of the electrode surface.
In this work, we use this SJ method to directly calculate elementary step energetics of the hydrogen evolution reaction on Pt(111) as functions of the applied potential.
Moreover, we relate these microscopic quantities to macroscopic behavior via microkinetic models and calculate Tafel slopes for all possible HER mechanisms. 

\subsection*{Methods}

We use the Solvated Jellium (SJ) method to incorporate potential into \textit{ab initio} electronic structure calculations.
This method, which has been published elsewhere \cite{Kastlunger2018}, allows for the addition/subtraction of electrons (or fractions thereof) to/from the unit cell in order to keep the electrode potential at a target value.
The SJ method is implemented in the GPAW \cite{GPAW1, GPAW2} electronic structure code and allows for systematic and selective charging of the metal-solvent interface.
This method naturally localizes the excess charge only at the top (reactive) side of the electrode, avoids spurious solvation effects, and solves the generalized Poisson equation to deduce the electrostatics; full details are provided in Kastlunger \textit{et al.}~\cite{Kastlunger2018}.
The dipole-induced electric field is screened with implicit and explicit water, where the implicit solvent module is a grid-based model developed by Held and Walter \cite{Held_implsolv_2014}.
The explicit water is modeled as a hexagonal, icelike structure with an H-down geometry (\textit{i.e.} every other water molecule has a hydrogen atom pointing towards the surface).
This interfacial water geometry has been shown to be the most stable geometry for protonated, that is, acidic water bilayers, at the potentials of interest for HER \cite{Schnur2009, Tonigold2012, Sakong2016, Kastlunger2018}.
The reaction pathways are all performed at constant potential within a tolerance of 0.01~V, and the relevant energy comparison is therefore the energy at constant potential (often referred to as $\Omega$~\cite{Alavi_JCP_2001, Sugino_PRB_2006, Neurock_2006, Jinnouchi2008, Letchworth-Weaver2012, Head-Gordon_2016,Sundararaman_GC_2017, Kastlunger2018, Bouzid2018, vdBossche2019, Melander2019}),

\begin{equation}
 E^\pot = E^{\nne} + \nne \,  \pe,
\label{eq:grand_pot}
\end{equation}

\noindent
where $E^{\nne}$ is the constant-charge (canonical) potential energy at an applied charge of -\nne, corresponding to $\pot$. The post-processing correction term includes the electrode work function, $\pe$ (in the article also noted as $-\mu^\phi_{\ce{e}^-}$), and the number of added/subtracted electrons \nne. In the complete free energy reaction diagrams presented in the main article, free energy contributions are included via normal mode analysis in the harmonic limit at 298 K \cite{Peterson2010}

\begin{equation}
 \Delta G^\pot = \Delta E^\pot + \Delta ZPE + \int_0^T     C_p \ dT - T \Delta S,
\label{eq:free}
\end{equation}

\noindent
where $\Delta G^\pot$ is the free energy at constant potential composed of $\Delta E^\pot$ (eq.~\ref{eq:grand_pot}) extrapolated to 0 K and vibrational free energy corrections calculated from normal mode analyses in a harmonic approximation. We observe one imaginary frequency at the transition state for each elementary reaction. This imaginary frequency represents an M-H or O-H stretch---the intuitive saddle point geometries for these reactions.

The Pt(111) electrode is modeled as a 3$\times$2$\times$3 slab, where the bottom layer is fixed at the optimized lattice constant (3.97 \AA{}). The systems are sampled on a 4x6x1 \textbf{k}-point mesh. The unit cell is 30 \AA{} in the \textit{z}-direction and periodic in \textit{xy}, and a dipole-correction \cite{Bengtsson1999} is included in the \textit{z}-direction. All calculations are performed at the generalized gradient approximation (GGA) level with the Perdew-Burke-Ernzerhof (PBE) \cite{Perdew1996} exchange-correlation functional. The reaction barriers are calculated using the climbing-image nudged elastic band (CI-NEB) \cite{Henkelman2000, CINEB} approach, and all systems were relaxed until the forces acting on each unconstrained atom were below 0.05 eV/\AA{} and 0.03 eV/\AA{} for kinetic barriers and stable endstates, respectively. The geometries at each applied potential are tabulated in an ASE-compatible dictionary in the supporting information. A schematic of a typical unit cell is shown Figure S12 in the supporting information.

\subsection*{Potential dependence of barriers}
Minimum energy pathways were calculated with the SJ method at a range of fixed voltages, shown in Figure~\ref{fig:bands}; the key figures relevant to the discussion are shown here, while additional scenarios are found in the supporting information.
We find that the barrier for the Volmer step is relatively low, but a strong function of potential.
Since approximately a monolayer of \hupd\ is expected to be present at the onset of hydrogen evolution, Figure~\ref{fig:bands} shows the barrier for adsorbing an on-top H while all hollow sites are saturated.
Barriers were also low for proton discharge to the same (on-top) site on pristine Pt(111), \textit{i.e.}, Pt(111) with vacant fcc-hollow sites~\cite{Kastlunger2018}. For proton discharge into the fcc-hollow sites, we observe that the reaction proceeds via an on-top site, followed by diffusion into the fcc-hollow site. This is shown in the supporting information.
As shown in Figure~\ref{fig:bands}, the reaction energy of this step changes linearly with applied potential, while the reaction barrier decreases quadratically with increasing overpotentials.

The Heyrovsky reaction---the electrochemical \ce{H2} liberation step---is also a strong function of potential, and Figure~\ref{fig:bands} shows that it is a much more difficult process than the Volmer step.
This figure shows the barrier involving an on-top H*; when we calculate the Heyrovsky reaction starting from a hollow-bound H*, we find the atom first diffuses to an on-top site before reacting, resulting in a higher barrier in net.
Compared to the Tafel reaction barriers, we see it is higher at most reasonable potentials, although a microkinetic model will shed greater insight into the relative pathways.

Functionally, the barriers of the Volmer and Heyrovsky steps exhibit similar, non-linear behavior with respect to potential.
As a simple model of this behavior, we can consider the initial and final states to lie in quadratic potential energy wells, and the reaction barrier to be described by the point of intersection, (functionally) similar to Marcus theory \cite{Marcus1956, Marcus1965}.
This form captures the basic physics expected~\cite{Hammond1955}: as the elementary reaction energy becomes strongly exothermic ($\Delta E \ll 0$), we expect the reaction to smoothly approach an activationless state ($E^\ddag \rightarrow 0$); similarly, when the elementary reaction becomes highly endothermic ($\Delta E \gg 0$), we expect no additional barrier beyond the energy change of the reaction ($E^\ddag \rightarrow \Delta E$).
This simple parabolic picture leads to the relation:
\begin{equation}\label{eq:marcus-type}
E^\ddag = \left\{
\begin{array}{lll}
0 &, & \Delta E < -4 b \\
\frac{(\Delta E + 4 b)^2}{16b} &, & -4b \le \Delta E \le 4b \\
\Delta E &, & \Delta E > 4b
\end{array}
\right.
\end{equation}

\begin{figure*}
  \includegraphics[width=0.48\textwidth]{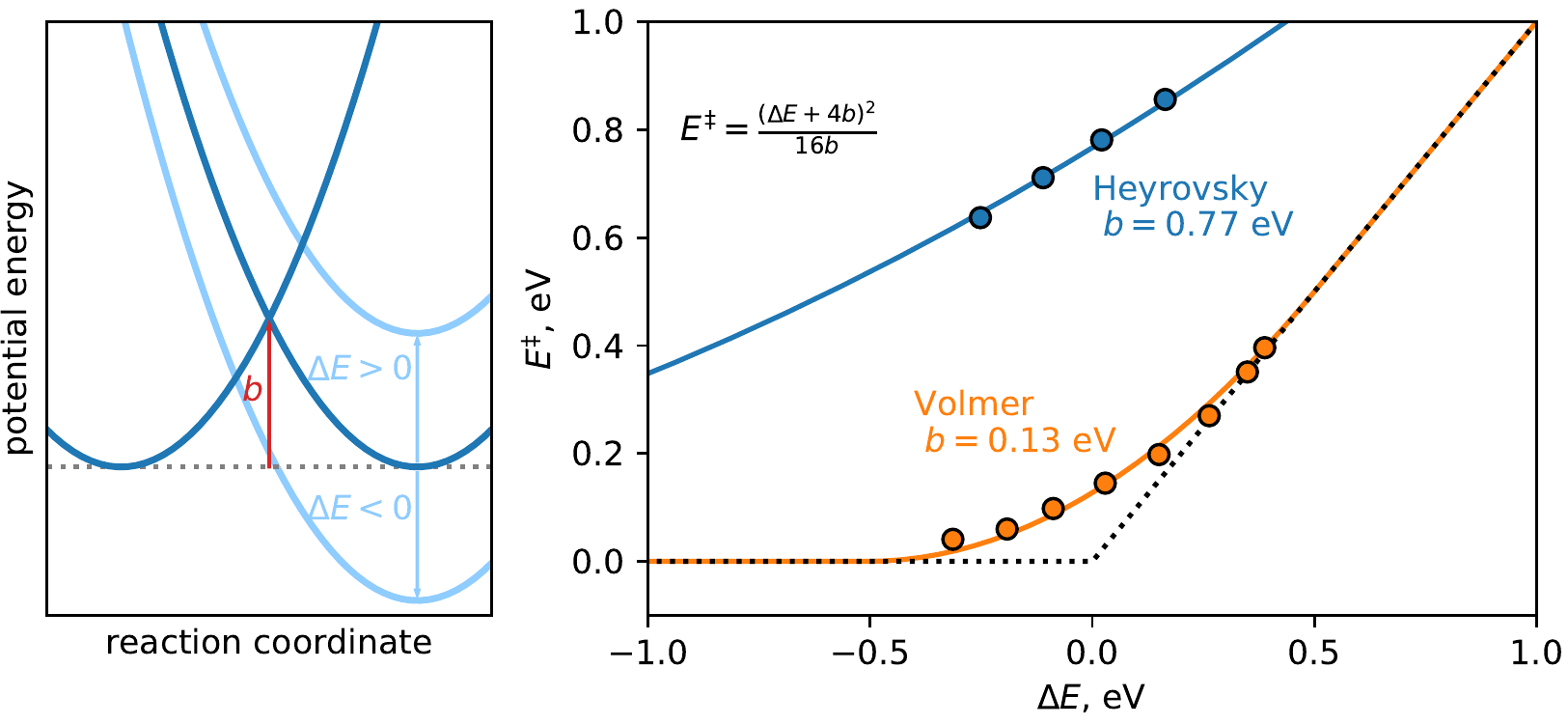}
	\caption{Reaction energies and barriers of the Volmer and Heyrovsky reactions fitted to the functional form in eq. (\ref{eq:marcus-type}). The relationship is quadratic for intermediate driving forces and linear for high-magnitude driving forces.}
  \label{fig:marcus}
\end{figure*}

\noindent
Here, we choose to formulate this with the parameter $b$, which can simply be interpreted as the barrier when the elementary step is thermoneutral ($\Delta E = 0$); that is, the intrinsic barrier.
The two potential-dependent steps are plotted in this form in Figure~\ref{fig:marcus}, where we can see this one-parameter model captures the data well.
The limiting behavior is evident for the Volmer reaction; the reaction approaches the activationless and barrierless regions for highly negative and positive potentials, respectively.
The difference in the reactions is quite apparent by the difference in the calculated intrinsic barriers: $b$ is only 0.13 eV for depositing a proton onto a top site (Volmer), but is 0.77 eV to liberate \ce{H2} electrochemically (Heyrovsky).
We assert that this one-parameter model is superior to both the two-parameter Br{\o}nsted-Evans-Polanyi (BEP) relations \cite{Bronsted1932, Evans1938, Bligaard2004} and linear transition-state scaling relations \cite{Alcala2003, Wang2011, Sutton2012} that are in common use for catalytic reactions, as the parabolic scaling smoothly captures the limiting forms and appropriate curvatures, making it ideal for use in microkinetic models.

The Tafel step---non-electrochemical \ce{H2} liberation---is not electrochemical in nature, and its barrier as calculated in the SJ method shows this: its energetics are essentially independent of the applied potential (Figure~\ref{fig:bands}).
The barriers show no potential dependence, and the slight potential dependence of the reaction energy is a consequence of a small change in surface dipole (due to reorientation of the water bilayer when \ce{H2} is incorporated); this energetic spread is negligible compared to the Volmer and Heyrovsky reactions.
A much stronger effect is the originating sites of the H atoms involved: the combination of two top-site bound hydrogens is much easier than that of a top and hollow H*.
We also calculated this for two hollow-adsorbed atoms, and found an even higher barrier, nearly 1~eV, as shown in the supporting information.
A comparison of the Tafel and Heyrovsky reactions shows that both the top--top and top--hollow Tafel barriers are considerably lower at the potentials of interest for HER.

\subsection*{Potential-controlled free energy diagrams}
\begin{figure*}
  \begin{center}
  \includegraphics[width=0.49\textwidth]{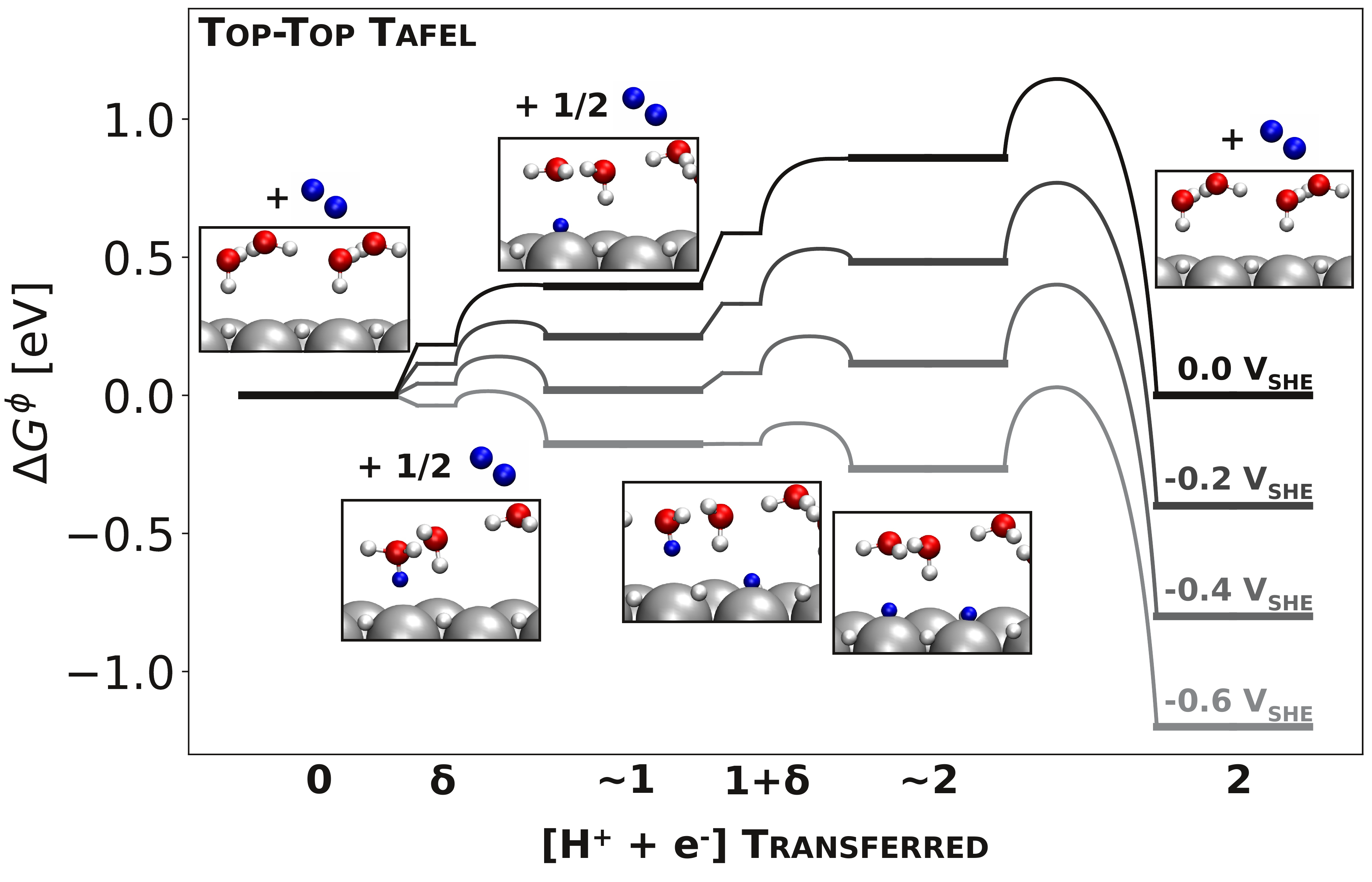}
   \caption{Free energy diagram for the top--top Volmer--Tafel mechanism at $\phi$ $\leq$ 0 V$_{\mathrm{SHE}}$.}
  \label{fig:FE_ttTafel}
 \end{center}
\end{figure*}

While there are studies that investigate the potential dependence of electrochemical reactions with grand canonical DFT methods~\cite{Head-Gordon_2016, Singh2017, vdBossche2019}, only a few have extended the analysis to ensure thermodynamic consistency~\cite{Steinmann2015}. Here we briefly describe how we accomplish thermodynamic consistency, while more complete details are in the SI.
The free-energy change between elementary steps separated by integer coupled proton--electron steps has been standardized via the computational hydrogen electrode (CHE) approach~\cite{Norscov_Ross_JPC_2004,Peterson2010}.
Here, we adapt this approach to non-integer, non-coupled portions of the reaction path, and unify the treatment of all portions of the reaction pathway.
First, we note that at a fixed potential $\phi$, an elementary step can deviate slightly from an integer electron transfer, as conventionally written.
That is, the Volmer step should properly be written as \ce{$*$ + H+ + (1 + \epsilon) e- -> H$*$}, and we modify the CHE equation to properly account for this:
\[ \dgrp = \mup{H$*$} - \mup{$*$} - \epsilon \mup{e-} - \frac{1}{2} \muo{H2} +  e \pot \]
\noindent
where the superscript $\phi$ indicates the chemical potential ($\mu$) is calculated at electrode potential \pot, and the term involving $\epsilon$ accounts for the non-integer charge transfer, which is corrected by the identical Fermi energy of the metal in both states.
The above expression effectively decouples proton and electron transfer, and allows for non-integer electron transfer.
This is precisely what we need to describe any position along the reaction path---including the transition state---which we can then express as

\[  G^\pot_\mathrm{path} - G^\pot_\mathrm{IS} = \mup{H-$*$} - \mup{$*$} + (1 - x) \mup{e-} - \frac{1}{2} \muo{H2} + e \pot \]

\noindent
where $\ce{H-$*$}$ is the reacting complex and $x$ is the (non-integer) number of electrons the circuit has to provide in order to drive the reaction. 
In this way, we automatically account for the shuttling of the proton from the bulk to the near-electrode surface, which can be accompanied by a partial charge transfer.
This fractional charge transfer has been investigated by Chen and coworkers \cite{Chen2018}, who found that the partial charge transfer is not a consequence of self-interaction in DFT~\cite{Perdew1981, Grafenstein2004, Toher2005, Stadler2012, Kastlunger2013}, but rather due to hybridization of the metal electrode and the protonated water bilayer when the additional proton is located close to the interface.
In our calculations, this phenomenon is evident from both Bader charge analysis \cite{Henkelman2006} and the grand-canonical description itself, where the latter indicates that only fractional numbers of electrons need to be injected into the system in order to keep to potential constant throughout the reaction.
On our free energy diagrams (\textit{e.g.}, Fig.~\ref{fig:FE_ttTafel}), the free-energy change of this proton-shuttling step is apparent as the shoulder before the barrier.
Note that this treatment is completely general for both electrochemical and non-electrochemical steps: we treat the Tafel step identically to the Volmer step, but find that its charge transfer is $\sim$0.
In this way, we do not decide \textit{a priori} what is an electrochemical step, but determine this \textit{a posteriori}---that is, by simply examining the resulting charge transfer.

Using this approach, we are able to create thermodynamically consistent free energy diagrams  for the competitive pathways in this reaction.
In Figure~\ref{fig:FE_ttTafel} we show the diagram for what these calculations suggest is the dominant reaction pathway: the \ce{H2} is liberated via the reaction of two top-bound H's in a non-electrochemical release.
Diagrams for the competitive pathways are included in the SI; the dominance of the top--top pathway is justified via the microkinetic models described in the following sections of this letter.

The mechanism begins with proton shuttling from the bulk into the metal--solvent interface followed by a Volmer step to form a top-bound H species.
The initial proton shuttling step is endergonic in free energy at low overpotentials; we interpret this as a consequence of the decreased entropy of the solvated proton in the confined electrochemical double layer relative to its bulk solvated counterpart; the entropic reduction, $\Delta (TS)$, is $-0.3$ to $-0.4$ eV at room temperature for all studied diffusion steps.
Perhaps surprisingly, the free energy diagram is starkly uphill, which is in contrast to the typical theoretical picture of the reaction proceeding through a $G{\sim}0$ hydrogen.

\subsection*{Microkinetic Models}
\begin{figure*}
  \begin{center}
  \includegraphics[width=0.49\textwidth]{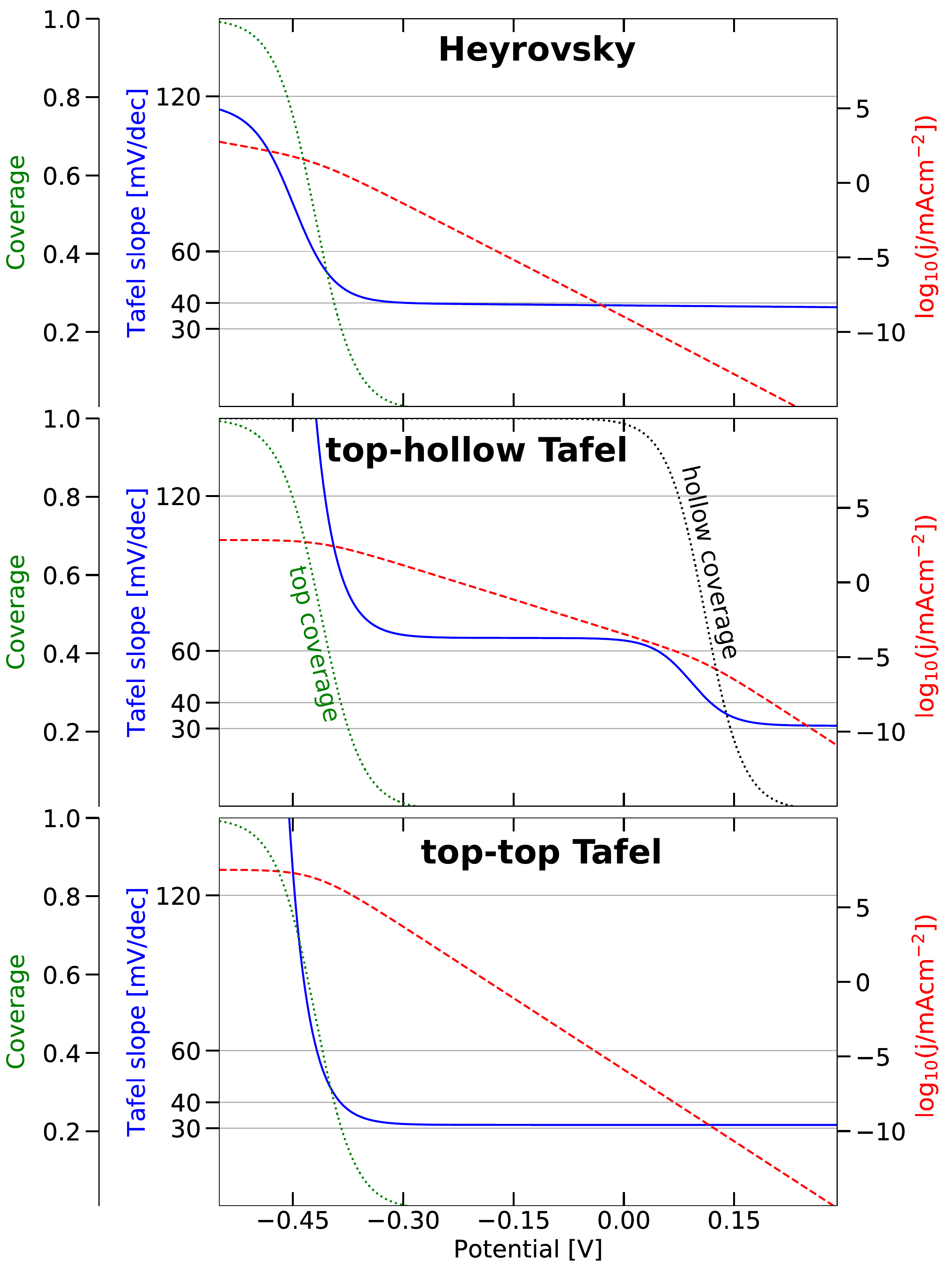}
   \caption{Coverage-dependent Tafel slopes (blue) and current densities (red) for the three reaction mechanisms considered. The coverages are shown in dotted lines. When applicable, we distinguish between top- and hollow-site coverages.}
  \label{fig:micro}
 \end{center}
\end{figure*}

We extend the atomistic results above to macroscopic observables via microkinetic models; the implementation details are included in the supporting information.
Figure~\ref{fig:micro} shows the potential-dependent current densities, Tafel slopes and coverages of the three studied mechanisms.
The names in the figure refer to the rate-limiting step; all three mechanisms also contain Volmer (proton deposition) steps, which will become important in the subsequent discussion. 
The Heyrovsky path has the lowest current over the reported range of potentials and appears to be of little importance except possibly at large overpotentials, discussed later.
The Tafel mechanism---limited by H$*$ combination---is the dominant pathway, which is considered for two cases.
The top--top mechanism dominates over the broadest range of potentials, although at potentials near the equilibrium potential the top--hollow Tafel may be significant.

Of particular importance in mechanistic studies of electrocatalytic systems are Tafel slopes ($d \phi / d \log_{10} j$), which reveal the relationship between driving force (potential) and reaction rate (current)~\cite{Fletcher2009, Shinagawa2015, Ledezma-Yanez2017}. 
We offer an intuitive picture of Tafel slopes for this study.
While the full microkinetic model presented in Figure~\ref{fig:micro} includes all barriers, let's instead assume that the hydrogen-liberating step has a rate equation of
\begin{equation}\label{eq:rkaa}
r = k^\ddag \, a_1 \, a_2
\end{equation}
\noindent
where $\{a_i\}$ indicate thermodynamic activities of the reactants.
As shown in the supporting information, this rate-limiting step can be expressed as
\[ r \propto
\exp \{ {-(\underbrace{\Delta G^\ddag + \Delta G_1 + \Delta G_2}_{\equiv \gnet})/k_\mathrm{B} T} \} \]
\noindent where $\Delta G^\ddag$ is the reaction barrier and $\{\Delta G_i \}$  are from equilibrated reactant activities when $a_i \ll 1$.
(When $a_i \sim 1$, the associated $\Delta G_i$ drops out of \gnet.)
Since $j \propto r$, the Tafel slope is $d\phi/d \log r = - k_\mathrm{B} T \ln 10 \cdot d\phi/d(\gnet) = d\phi/d( \gnet ) \cdot 59$ meV/dec (at 298~K).
That is, the Tafel slope can be directly inferred from the potential dependence of $\gnet$.

Consider first if the Tafel step (\ce{H$*$ + H$*$}) is rate-limiting: \gdag\ is not a function of $\phi$, so the only potential dependence comes from $a_1 a_2$, both of which are coverages.
When coverages are low ($\theta \ll 1$), $\Delta G_\mathrm{net} = \Delta G_1 + \Delta G_2 \approx 2e\phi$, and $d\phi/d(\Delta G_\mathrm{net}) = 1/2$, giving a Tafel slope of 30 mV/dec.
This matches both the top--top and top--hollow behavior in Figure~\ref{fig:micro} when coverages are low. 
However, when a coverage approaches unity, its activity in equation~\eqref{eq:rkaa} becomes a constant and a factor of $\Delta G_i$ drops out of $\Delta G_\mathrm{net}$, giving a slope of 59 meV/dec, as seen in the top--hollow mechanism after the hollow sites saturate.
By identical logic, when both coverages approach unity, the rate becomes insensitive to voltage and the Tafel slope $\rightarrow \infty$; we see this at large negative potentials for both top--top and top--hollow.  This behavior a priori rules out the hollow--hollow mechanism, which has been proposed in earlier theoretical studies, since the UPD coverage approaches unity at 0 \vshe\ in both the results presented here and experimental studies~\cite{Ogasawara1994, Markovic1996, Morin1996, Conway1998, Herrero2001, osiewicz2012}.
In summary, the rate changes rapidly when coverages change rapidly, which can only occur when coverages are very low; that is, orders-of-magnitude below 1, suggesting that only two weakly-bound H$*$'s can be responsible for a Tafel slope of 30 mV/dec.

For the Heyrovsky mechanism, the reaction barrier is a function of $\phi$ as we calculated in Figures~\ref{fig:bands}--\ref{fig:marcus}; by linearizing eq.~\eqref{eq:marcus-type} (about zero driving force) we directly estimate $\gdag = \Delta G^{\ddag}_0 + \beta e\phi$, with $\beta=0.5$.
$a_1$ is the activity of a dissolved proton, which is constant at fixed pH, so $\Delta G_1$ is absent from \gnet.
$a_2$ is again hydrogen coverage.
Therefore, when $\theta \ll 1$, $\Delta G_\mathrm{net} = 1.5 e \phi$ and the Tafel slope is 39 mV/dec, matching that seen in the microkinetic model.
When $\theta \approx 1$, $\Delta G_\mathrm{net} = 0.5 e\phi$ and the Tafel slope is 118 mV/dec, matching that seen microkinetically when sites saturate.

Based on this simple analysis, we can conclude that the only way to achieve a Tafel slope of $\sim$30 mV/dec is in the limit when both activities correspond to coverages that are changing exponentially.
In our analysis, this can only be caused by the weak-binding hydrogens (top--top), and the UPD hydrogen must be a kinetic spectator over most HER potentials.

While the Tafel slopes are in excellent agreement with experimental results, the current densities are not.
This is to be expected, for a number of reasons.
First, in the absence of an internal reference potential, we use an approximate value of 4.4 V for the absolute standard hydrogen electrode potential; current responds exponentially to potential.
Second, it can be challenging to accurately assign the unity activity reference state from limited-size atomistic calculations, affecting pre-factors in transition-state theory.
Another source of uncertainty is the \hupd{} coverage; experimental studies have disagreed as to whether the \hupd{} coverage is complete or fractional (0.67 ML) at the equilibrium potential on Pt(111).
This is often attributed to site-blocking effects, since anion groups must desorb before the fcc sites are saturated with hydrogen~\cite{Conway2000}.
Since we do not explicitly include the electrolyte, save for an explicit water layer, this effect is not accounted for in the simulations.
Finally, DFT itself has limited accuracy, and deviations of 0.1--0.2 eV are to be expected.
The quadratic dependence of the reaction rate on the coverage makes the reaction rate highly sensitive to the reaction energy of the coverage-determining step (Volmer).
A change in reaction energy of the Volmer step by 0.1 eV changes the overpotential to reach -10 mA/cm$^2$ by 100 mV; this is well within the error of DFT. 
Moreover, we note that a slight cathodic shift of the reference potential by $\sim$150 mV leads to experimentally observed exchange current densities in combination with a Tafel slope of 30 mV dec$^{-1}$.
Interestingly, at nearly the same potentials the top-top Tafel mechanism becomes the dominating mechanism and supersedes the top-hollow mechanism, which exhibits higher currents at more anodic potentials. 

\subsection*{Why Pt excels}
Although many further refinements can be made to such a model, we now have enough information to understand why platinum excels at the hydrogen evolution reaction.
The prevailing view of Pt's high reactivity is that the hydrogen binding (free) energy is near zero.
Simultaneously, much speculation has persisted on the role of the UPD hydrogen atoms: are they kinetically active or merely spectators?
However, our analysis makes clear that there must be \emph{two} exponentially varying coverages as driving forces in order to see such high current response to potential (30 mV/dec): this can only occur with hydrogen atoms that bind significantly weaker than $\Delta G = 0$; that is, the top-site hydrogens must be active.
This is in agreement with experimental conclusions reached by He and colleagues \cite{He2017}.
The \gz\ hydrogens are then kinetically inactive---but do promote the kinetic response by allowing for the weaker on-top hydrogens to become kinetically accessible.
The weakly-bound hydrogens have both exponentially-varying coverage and a low barrier for the non-electrochemical Tafel step.

\begin{figure*}
  \begin{center}
  \includegraphics[width=0.8\textwidth]{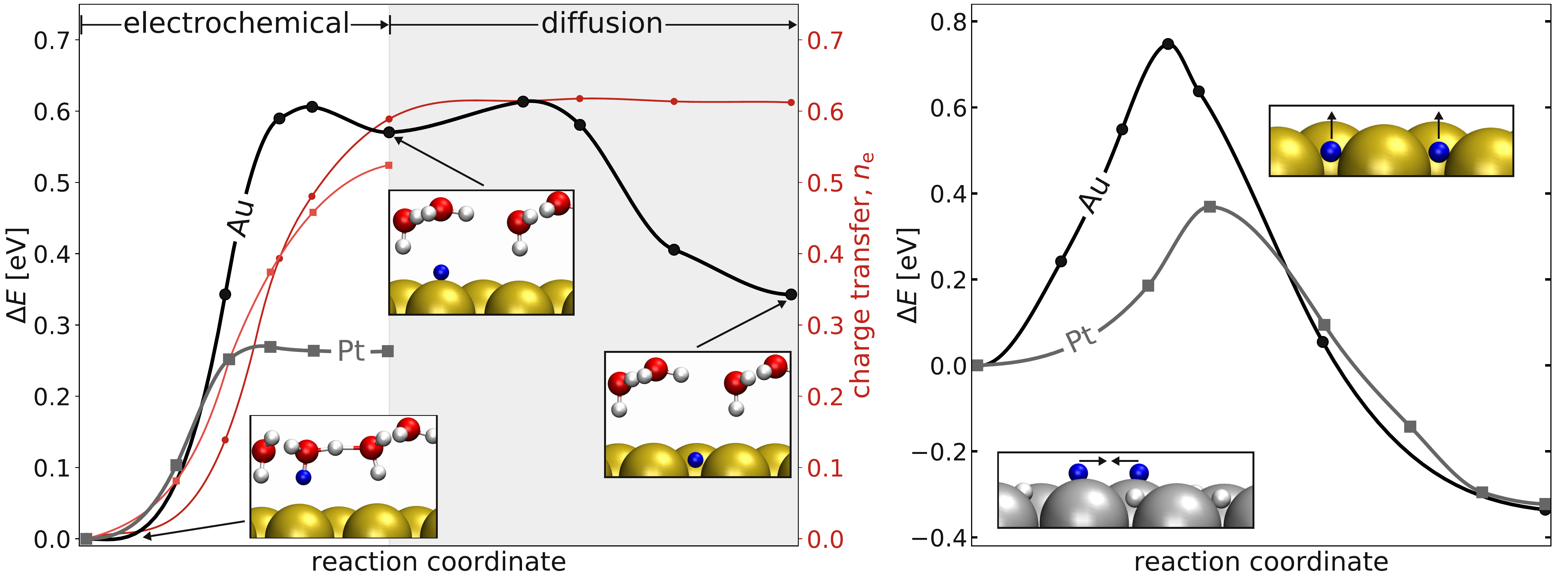}
  \caption{
  (Left) Minimum energy pathways (black) and charge transfer (red) curves  for the Volmer step on Pt and Au  at the equilibrium potential. The electrochemical reaction is essentially barrierless on both metals, but Au includes an additional surface diffusion step from an on-top site to a neighboring fcc-hollow site. (Right) \ce{H2} liberation on Au and Pt via a Tafel step. Although the energetics are nearly identical, Au has a much higher barrier due to the hollow, rather than top, site.}
  \label{fig:ec_diff}
 \end{center}
\end{figure*}

This leads to an obvious question: if the role of Pt's \gz\ hydrogen is to allow weaker hydrogen binding, why wouldn't more noble metals, such as gold, be highly active as well?
On such a noble system, we'd expect to have exponentially-varying coverages, and transition-state scaling suggests that we would also have low barriers. 
However, in Figure~\ref{fig:ec_diff} we show that despite similar hydrogen-binding energies, both the Volmer and Tafel barriers on Au are significantly higher than on Pt.
This is a consequence of the reactions proceeding from \emph{hollow} sites on Au, rather than top sites on Pt.
We first examine the \ce{H2}-liberating Tafel reaction which limits the rate on Pt; this is not an electrochemical reaction.
The energetics of this reaction on Au and Pt are nearly identical; however, the barrier on Au is higher by nearly a factor of 2.
The reaction from the hollow sites on Au follows a high-energy stretch which is predominantly perpendicular to the surface; while the reaction from Pt's top sites follows lower-energy lateral bending modes; we quantitatively project the reaction path onto these normal modes in the SI.
We also compare this barrier on Pt to several other surfaces, where we see that the top site has a much lower barrier that deviates from the traditional scaling relation for this reaction.
For the proton deposition (Volmer) step on Au, we see the overall energetics are very similar to Pt, but the barrier is much higher on Au.
We can understand this by examining the reaction pathway and electron-transfer characteristics: to deposit a hydrogen in a hollow site on Au, the reaction first proceeds through an even more weakly bound top site, before diffusing into the stable hollow site.
The diffusive portion of the reaction is not electrochemical, as evidenced by the lack of charge transfer, and thus does not change with applied voltage.
Hence, the electrochemical reaction on both metals displays similar characteristics (\textit{i.e.}, symmetry factor close to unity), but the intrinsic reaction barrier on Au(111) is significantly higher.
Here, this raises the effective barrier on Au from essentially no ``extra'' barrier (beyond $\Delta E$) on Pt, to $\sim$0.35~eV extra barrier on Au.
We have verified that this reaction pathway is not an artifact of the hexagonal water structure on Au by also calculating the pathway from a proton dimer (\ce{H5O2+}), as discussed in the SI.
Both our calculations (see supplementary material) and experimental studies~\cite{Kahyarian2017, Khanova2011} suggest that the increased barrier of the Volmer step causes it to become rate-limiting for Au surfaces.
Thus, although Au's reactive hydrogens have nearly identical binding energies to Pt, the hollow sites available on Au have intrinsically more challenging reaction paths, which rationalizes the much higher reactivity of Pt at similar binding energies.

Thus, the role of the \gz\ hydrogens in Pt---which are experimentally observed as a capacitative event just before the onset of HER---is to \emph{both} weaken the hydrogen bonding strength \emph{and} allow reactions to proceed through a more facile, on-top, mechanism.
Thus, we can understand why Pt is so uniquely reactive: more noble metals are forced into a hollow--hollow mechanism that carries kinetic penalties, while more reactive metals---which can employ the top--top mechanism---will also have higher Tafel barriers (via scaling) and further would bind hydrogen too strongly to have exponentially variable coverages, giving sluggish current responses.
We hope that these insights can lead to the atomic-scale design of alternative catalysts with activities that rival Pt.

\vspace{5mm}

We gratefully acknowledge funding from the Office of Naval Research, award N00014-16-1-2355. We would also like to thank Tim Mueller, Chenyang Li, Peter Weber and Maxime Van den Bossche for their valuable input.
High-performance computational work was carried out at the Center for Computation \& Visualization (CCV) at Brown University.

\subsection*{Supporting information}
The supporting information contains a complete derivation of the decoupled computational electrode, free energy diagrams of competing reaction pathways, tabulated energies, charge transfer profiles as a function of the applied potential, microkinetic implementation details, Tafel slope derivations, additional examples and geometries for all structures.

\end{multicols}

\bibliography{her_mech_paper}
\newpage

\begin{figure*}
 \begin{center}
  \includegraphics[width=3.25in, height=1.75in]{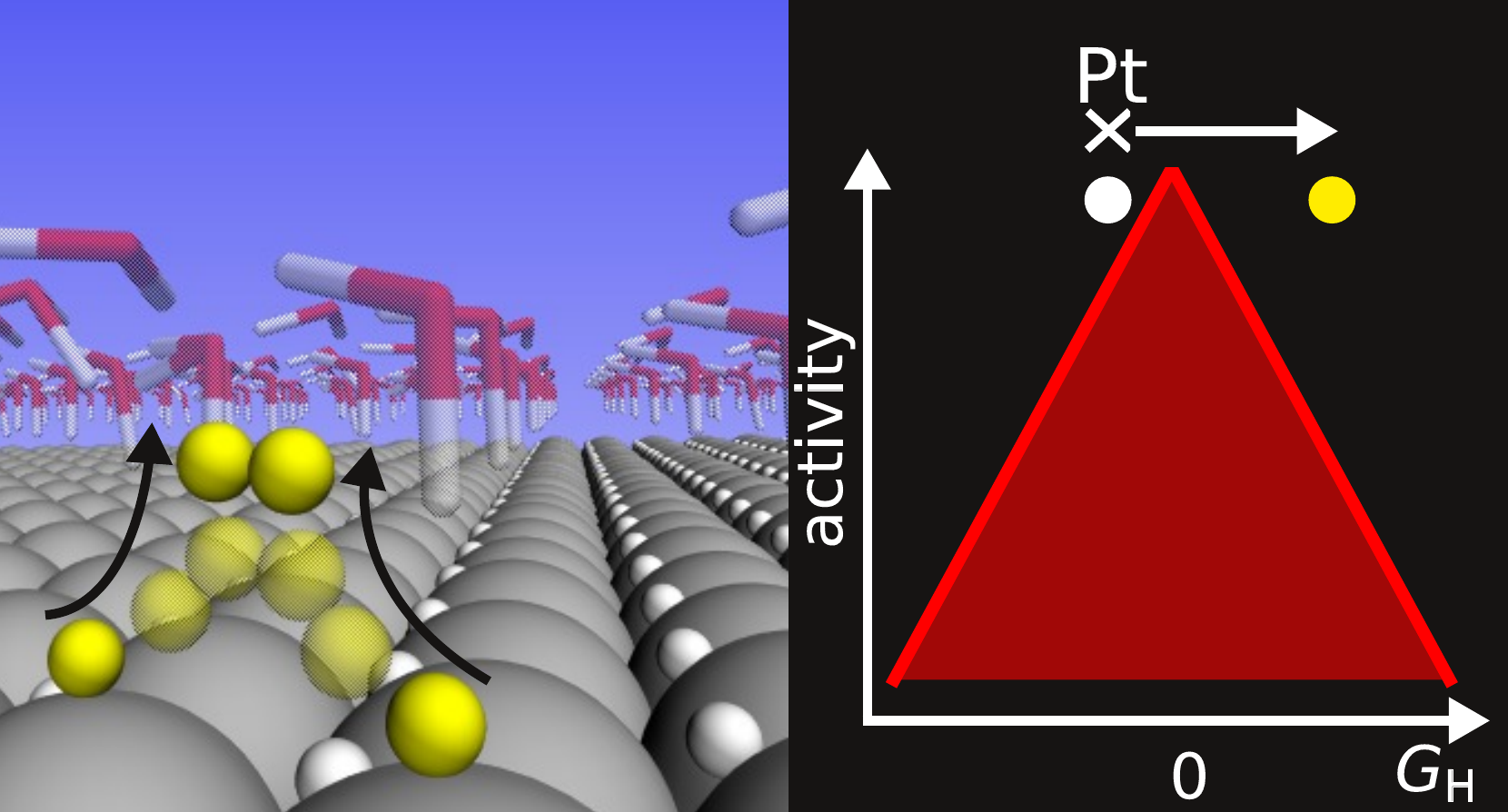}
   \caption*{For Table of Contents Only}
 \end{center}
\end{figure*}

\end{document}